\documentclass[conference]{IEEEtran} 

\usepackage{hyperref}
\usepackage{graphicx}
\usepackage{xspace}
\usepackage{todonotes}
\usepackage{subcaption}

\usepackage{pifont}

\newcommand{\SAP}{SAP\xspace}
\newcommand{\VULAS}{Vulas\xspace}
\newcommand{\code}[1]{\textsf{\footnotesize#1}\xspace}
\newcommand{\cve}[1]{\textsf{\footnotesize#1}\xspace}

\newcommand{\vulasprojectcount}{496\xspace}
\newcommand{\vulasprojectcountapprox}{500\xspace}

\newcommand{\vulasnumberofscansapprox}{250000\xspace}
\newcommand{\squeeze}[1]{\vspace*{-#1mm}}
\renewcommand{\squeeze}[1]{}

\newcommand{\note}[1]{{\textsf{\color{blue}{#1}}} \addcontentsline{tdo}{todo}{#1}}

\newcommand{\review}[1]{{\color{blue}{#1}}} 
\renewcommand{\note}[1]{}
\renewcommand{\review}[1]{#1}

\title{\LARGE \bf
Beyond Metadata: Code-centric and Usage-based Analysis \\of Known Vulnerabilities in Open-source Software \\
}

\author{\IEEEauthorblockN{Serena Elisa Ponta\IEEEauthorrefmark{1},
Henrik Plate\IEEEauthorrefmark{2} and
Antonino Sabetta\IEEEauthorrefmark{3}}
\IEEEauthorblockA{SAP Security Research,
Mougins, France\\
Email: \IEEEauthorrefmark{1}serena.ponta@sap.com,
\IEEEauthorrefmark{2}henrik.plate@sap.com,
\IEEEauthorrefmark{3}antonino.sabetta@sap.com}}

\begin{document}

\begin{figure*}
\begin{minipage}{\textwidth}

\begin{center}
  {\Large \bf Beyond Metadata: Code-centric and Usage-based Analysis \\[3mm] of Known Vulnerabilities in Open-source Software}
  \\[4mm]
  {\large[PRE-PRINT]}
  \\[8mm]
  {\large Serena Elisa Ponta, Henrik Plate, Antonino Sabetta}
\end{center}

\vspace{10mm}
\noindent\textsc{Abstract}
\input{abstract}
\vspace{15mm}
\hrule
\vspace{10mm}
\begin{center}
{\Large Citing this paper}
\end{center}

This is a pre-print of the paper that appears in the proceedings of the
34th IEEE International Conference on Software Maintenance and Evolution
(ICSME) 2018.

If you wish to cite this work, please refer to it as follows:

\begin{verbatim}
@INPROCEEDINGS{ponta2018icsme,
  author={Serena Elisa Ponta and Henrik Plate and Antonino Sabetta},
  booktitle={2018 IEEE International Conference
             on Software Maintenance and Evolution (ICSME)},
  title={Beyond Metadata: Code-centric and Usage-based Analysis
         of Known Vulnerabilities in Open-source Software},
  year={2018}
  month={Sept},
}
\end{verbatim}
\vspace{5mm}
\hrule

\end{minipage}
\end{figure*}

\maketitle

\begin{abstract}

The use of open-source software (OSS) is ever-increasing, and so is the number of open-source vulnerabilities being discovered and publicly disclosed. The gains obtained from the reuse of community-developed libraries may be offset by the cost of \review{timely detecting, assessing, and mitigating their vulnerabilities.}


In this paper we present a novel method to detect, assess and mitigate OSS vulnerabilities that improves on state-of-the-art approaches, which commonly depend on metadata to identify vulnerable OSS dependencies. Our solution instead is code-centric and combines static and dynamic analysis to determine the reachability of the vulnerable portion of libraries used (\emph{directly} or \emph{transitively}) by an application. Taking this usage into account, our approach then supports developers in choosing among the existing non-vulnerable library versions.

\VULAS, the tool implementing our \emph{code-centric} and \emph{usage-based} approach, is officially recommended by \SAP to scan its Java software, and has been successfully used to perform more than \vulasnumberofscansapprox~scans of about \vulasprojectcountapprox~applications since December 2016. We report on our experience and on the lessons we learned when maturing the tool from a research prototype to an industrial-grade solution.

\end{abstract}

\section{Introduction}

Open-source software (OSS) libraries are widely used in the software industry: by some estimates, as much as 80\% to
90\% of the software products on the market include some OSS component~\cite{snyk-oss-state}, and each of them contains,
on average, 100 distinct open source components, whose code weighs as much as 35\% of the overall application
size~\cite{blackduck2016}. The same study reports that for applications developed for internal use, the proportion is as
high as 75\%.

At the same time, the number of vulnerabilities disclosed for OSS libraries has been steadily increasing since
2009~\cite{snyk-oss-state}. While \emph{using OSS components with known vulnerabilities} is included in the \emph{OWASP
Top 10 Application Security Risks} since 2013~\cite{owasp-top-ten-2013,owasp-top-ten-2017}, still today the problem is
far from being solved. On the contrary, OSS vulnerabilities have been hitting the headlines of mainstream media many
times over the past few years~\cite{heartbleed,shellshock}. As reported by~\cite{snyk-databreach}, OSS vulnerabilities
were the root cause of the majority of the data breaches that happened in 2016. In 2017, the \emph{Equifax}
incident~\cite{equifax}, caused by a missed update of a widely used OSS component, compromised the personal data of over
140 millions of U.S. citizens.

Establishing effective OSS vulnerability management practices, supported by adequate tools, is broadly understood as a
priority in the software industry, and tools helping to \emph{detect} known vulnerable libraries are available nowadays,
either as OSS or as commercial products (e.g.,~\cite{owasp-dc,whitesource}).

These tools differ in terms of detection capabilities, but (to the best of our knowledge) the approaches they use rely
on the assumption that the metadata associated to OSS libraries (e.g., name, version), and to vulnerability descriptions
(e.g., technical details, list of affected components) are always \emph{available} and \emph{accurate}. Unfortunately,
these metadata, which are used to map each library onto a list of known vulnerabilities that affect it, are often
incomplete, inconsistent, or missing altogether. Therefore, the tools that rely on them may fail to detect
vulnerabilities (false negatives), or they may report as vulnerable artifacts that do not contain the code that is the
actual cause of the vulnerability (false positives).

Furthermore, merely detecting the inclusion of vulnerable libraries does not cater for the needs of the entire software
development life-cycle. In the early phases of development, updating a library to a more recent release is relatively
unproblematic, because the necessary adaptations in the application code can be performed as part of the normal
development activities. On the other hand, as soon as a project gets closer to the date of release to customers, and
during the entire operational lifetime, all updates need to be carefully pondered, because they can impact the release
schedule, require additional effort, cause system downtime, or introduce new defects.

To evaluate precisely the need and the urgency of a library update, it is necessary to answer the key question: ``is the vulnerability \emph{exploitable}, given the particular way the library is used within the application?''. Answering this question is extremely difficult: vulnerabilities are typically described in advisories that consist of short, high-level, textual descriptions in natural language, whereas a reliable assessment of the exploitability and the potential impact of a vulnerability demands much lower-level, detailed, technical information.


\review{Our previous work already} goes beyond the simple detection of OSS vulnerabilities:  the
approach we proposed in~\cite{icsme2015} analyzes the code changes introduced by security fixes, and uses dynamic analysis to assess the impact of the
vulnerability for a given application.


In this paper we build on \review{~\cite{icsme2015}}, proposing a \emph{code-centric} and \emph{usage-based} approach to \emph{detect}, \emph{analyze}, and \emph{mitigate} OSS vulnerabilities: \textbf{A)}
We generalize the vulnerability detection approach of~\cite{icsme2015} by considering fixes independently of the vulnerable libraries;
\textbf{B)} We use static analysis to determine whether vulnerable code is reachable and through which call paths; \textbf{C)} We \emph{combine} static and dynamic analysis to  overcome their mutual limitations;
\textbf{D)} We define metrics which support the choice of alternative library versions that are not vulnerable, highlighting which options are more likely to minimize the update effort
and the risk of incompatibility.
%

Our approach is implemented as a tool, \VULAS, which is adopted at \SAP as the officially recommended solution to scan
Java software. The tool has been successfully used to scan about \vulasprojectcountapprox~applications. Vulnerable code
was found reachable for 131 of them and we found that in 7.9\% of the cases this was only determined by the combination
of static and dynamic analysis. We report on our experience and on the lessons we learned when maturing \VULAS from a
research prototype to an industrial-grade solution that has been used to perform over \vulasnumberofscansapprox~scans
since December 2016.

The remainder of the paper is organized as follows: Section~\ref{sec:approach} describes the technical approach,
Section~\ref{sec:metrics} defines the update metrics, and Section~\ref{sec:evaluation} illustrates our approach in
practice. In Section~\ref{sec:experience}, we report on our experience, lessons learned and the challenges we
identified. Section~\ref{sec:rel-work} discusses related literature and Section~\ref{sec:conclusion} concludes the
paper.





\section{Technical description of the approach}
\label{sec:approach}

\review{In~\cite{icsme2015} we already presented} the idea of shifting the problem of establishing whether an
application is \emph{exploitable} because of a known vulnerability in an OSS library, to the problem of assessing
whether the vulnerable code is \emph{reachable}.

Sections~\ref{subsec:background},~\ref{subsec:id}, and~\ref{subsec:dynamic} \emph{generalize}~\cite{icsme2015}, whereas
Sections~\ref{subsec:static},\ref{subsec:combination} \emph{extend} it with unique novel contributions, that are the
basis of the update metrics presented in Section~\ref{sec:metrics}.





\subsection{Background}
\label{subsec:background}
The core of~\cite{icsme2015} lies on the assumption that a vulnerability can be detected and
analyzed considering the set of program constructs (such as methods), that were modified, added, or deleted to fix that vulnerability.




We define a \emph{program construct} (or simply \emph{construct}) as a structural element of the source code characterized by a \emph{type} (e.g., \code{package, class, constructor, method}), a \emph{language} (e.g., Java, Python), and a \emph{unique identifier} (e.g., the fully-qualified name\footnote{As an example, \code{foo.Bar.baz(int)} is the fully-qualified name of method \code{baz(int)} in class \code{Bar} and package \code{foo}.}).

\review{Changes to program constructs are done by means of \emph{commits} to a source code repository. The set of changes that fix a vulnerability can be obtained from the analysis of the corresponding commit,
the \emph{fix commit}}\footnote{In case a commit includes not only a vulnerability fix but also unrelated changes, then a post-processing of the construct changes is required.}.
We define a \emph{construct change} as the tuple
$$(c, t, \mathtt{AST}_f^{(c)},\mathtt{AST}_v^{(c)})$$
where $c$ is a  construct, $t$ is a change operation (i.e., addition, deletion or modification) on the construct $c$, and $\mathtt{AST}_f^{(c)}, \mathtt{AST}_v^{(c)}$ are, respectively, the abstract syntax trees of $c$ at commit $n$ and at commit $n-1$, i.e., the AST of the fixed and vulnerable construct. Notice that for deleted (added) constructs only $\mathtt{AST}_v^{(c)}$ ($\mathtt{AST}_f^{(c)}$) exists.


When a fix is implemented over multiple commits, we rely on commit timestamps to compute the set of changes by comparing the source code of the first and the last commit.
If the vulnerability fix includes changes in a  nested construct (e.g., a method of a class), two distinct entries are included in the set of construct changes, one for the outer construct (the class), one for the nested construct (the method).


\note{to revise if time allows}
The fix commits of a vulnerability are not always communicated with its disclosure. Some OSS projects (e.g., \code{Apache Tomcat}) provide such information via security advisories; others reference issue tracking systems which in turn describe the vulnerability being solved; some other OSS projects do not explicitly refer to vulnerabilities being fixed. Thus reconciling the information based on the textual description and code changes requires considerable manual effort \review{(see Section~\ref{subsubsec:db})}.
A broader discussion of the data integration problem can be found in~\cite{icsme2015}.



Differently from~\cite{icsme2015}, we provide a definition of construct change and consider the ASTs of the modified program constructs. This is used in Section~\ref{subsec:id} to establish whether libraries include the changes introduced by the fix.

\subsection{Vulnerability Detection}
\label{subsec:id}

\begin{figure}[t]
\centering
\includegraphics[width=\columnwidth]{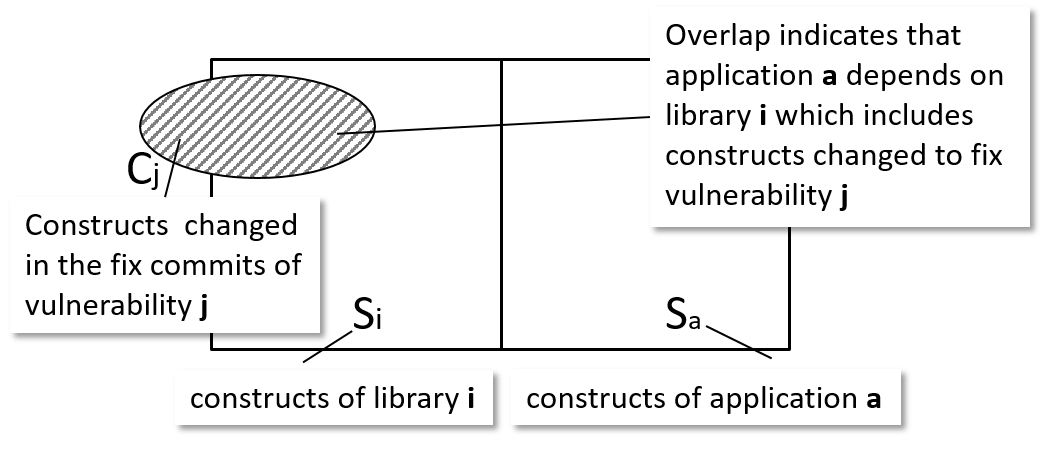}
\caption{Vulnerability detection}
\vspace*{-4mm}
\label{fig:concept-app}
\end{figure}

Figure~\ref{fig:concept-app} illustrates how a vulnerability $j$ is associated to an application $a$.
$C_{j}$ is the set of the constructs obtained as described above
by analyzing the fix commits of $j$.
The set $S_i$ is the set of all constructs of the OSS
library $i$ bundled in the application $a$ whereas $S_a$ is the set of
all constructs belonging to the application itself.



If $C_{j} \cap S_i \neq\emptyset$ and $\forall c \in C_{j} \cap S_i, \mathtt{AST}^{(c)} = \mathtt{AST}_v^{(c)}$,  then we
conclude that the application includes a library $i$ with code vulnerable to $j$ (referred to as \emph{vulnerable constructs}), i.e., constructs that have been changed in the fix commits of $j$. 


If $\exists \ c \in C_{j} \cap S_i \ | \ \mathtt{AST}^{(c)} \neq \mathtt{AST}_v^{(c)}$, then we relax the equality constraint and use both $\mathtt{AST}_v^{(c)}$ and $\mathtt{AST}_f^{(c)}$ to establish to which of the two $\mathtt{AST}^{(c)}$ is ``closest''. To this end, we use tree differencing algorithms~\cite{gumtree,changedistiller} and a library comparison method that we devised (whose description is omitted from this paper because of space constraints).
%
%
Manual inspection is still required whenever no automated conclusion can be taken, e.g., when $\exists \ c_1,c_2 \in C_{j} \cap S_i \; | \; \mathtt{AST}^{(c_1)} = \mathtt{AST}_v^{(c_1)}$ and  $\mathtt{AST}^{(c_2)} = \mathtt{AST}_f^{(c_2)}$.

Note that, even when a vulnerability is fixed by adding new methods to an existing class,
the intersection $C_{j} \cap S_i$ is not empty because, as explained in the previous section,
it would contain the construct for the class. If the fix includes the addition of a class, we assume that existing code is modified to invoke the new construct.

Differently from \cite{icsme2015},
we define the set of constructs $C_j$  as independent of any library $i$, and a vulnerability in a library is then detected through the intersection of its constructs with $C_j$. This approach has several advantages: First, it makes it explicit that the vulnerable constructs responsible for a vulnerability $j$ can be contained in any library $i$, hence, the approach is robust against the prominent practice of repackaging the code of OSS libraries. Second, it is sufficient that a library includes a subset of the vulnerable constructs for the vulnerability to be detected. Last, it improves the accuracy compared to approaches based on metadata, which typically flag entire open-source projects as affected, even if projects release functionalities as part of different libraries. \code{Apache POI}~\cite{poi}, for instance, is developed within a single source code repository but released as multiple libraries, each offering functionalities for manipulating different types of Microsoft Office documents.


Moreover, \review{in \cite{icsme2015} we focused} on newly-disclosed open-source vulnerabilities, and thus \review{could} assume that, at the time of disclosure, every library that includes constructs changed in the fix commit is vulnerable.
%
While the assumption holds at that moment in time,
it is not valid for old vulnerabilities.
In this case, one has to establish whether a given library contains the fixed code, which we support by comparing the AST of constructs in use with those of the \emph{affected} and \emph{fixed} construct versions.


\subsection{Dynamic Assessment of Vulnerable Code}
\label{subsec:dynamic}
After having determined that an application depends on a library that \emph{includes} vulnerable constructs, it is important to establish whether these constructs are \emph{reachable}. \review{In this paper, we use the term \emph{reachable} to denote both the case where dynamic analysis shows that a construct is \emph{actually executed} and the case where static analysis shows \emph{potential} execution paths.}
The underlying idea is that if an application executes (or may execute) vulnerable constructs, there exists a significant risk that the vulnerability can be exploited.
The dynamic assessment described here is borrowed from~\cite{icsme2015}.


\begin{figure}
    \centering
    \begin{subfigure}[b]{0.45\textwidth}
        \includegraphics[width=\textwidth]{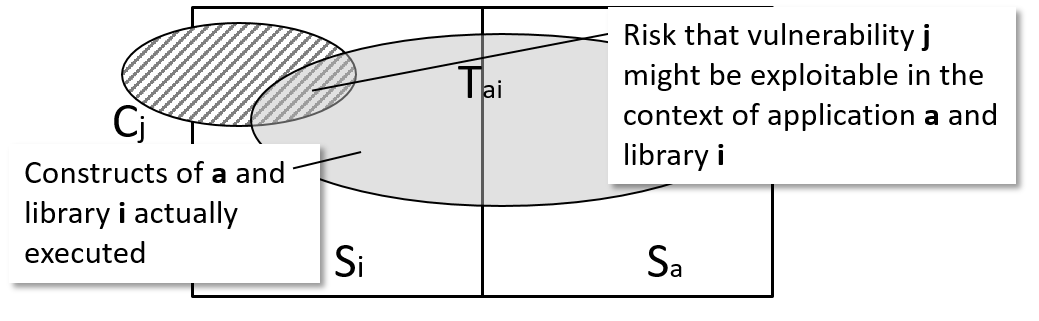}
        \caption{Dynamic}
        \label{fig:concept-test}
    \end{subfigure}
    ~ 
    \begin{subfigure}[b]{0.45\textwidth}
        \includegraphics[width=\textwidth]{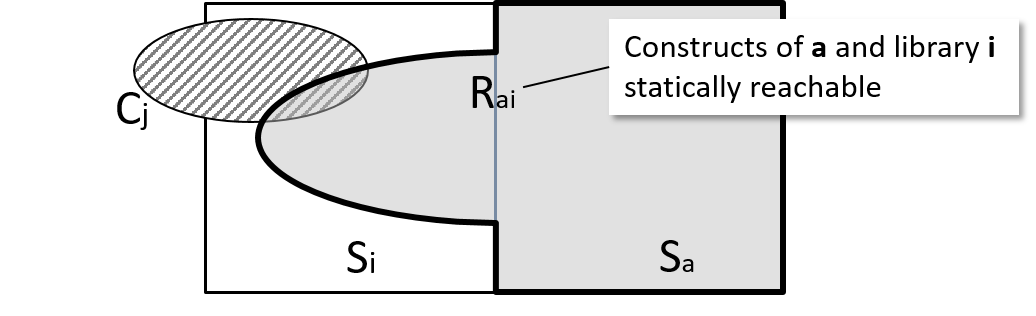}
        \caption{Static}
        \label{fig:concept-a2c}
    \end{subfigure}
    ~ 
    \begin{subfigure}[b]{0.45\textwidth}
        \includegraphics[width=\textwidth]{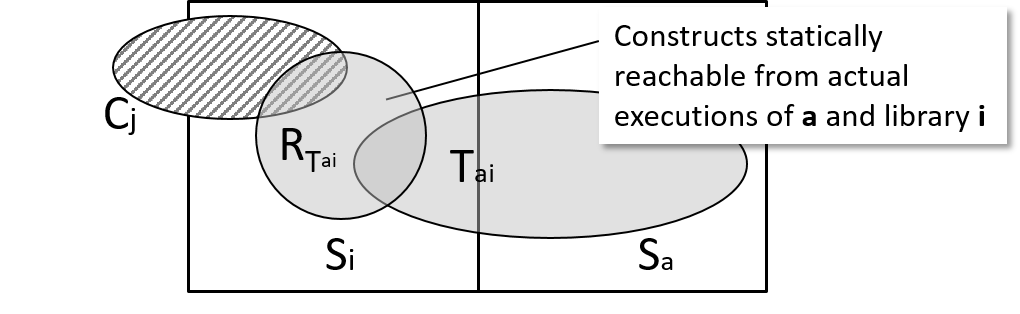}
        \caption{Combination of static and dynamic}
        \label{fig:concept-t2c}
    \end{subfigure}
    \caption{Vulnerability analysis}\label{fig:analysis}
     \squeeze{4}
\end{figure}

Figure~\ref{fig:concept-test} illustrates the use of dynamic analysis to assess whether the vulnerable constructs are reachable by observing actual executions. $T_{ai}$ represents the set of constructs, either part of application $a$ or its bundled library $i$, that
were executed at least once during some execution of the application.
The collection of actual executions of constructs can be done at different times: during unit tests, integration tests, and even during live system operation (if possible).
The intersection $C_{j} \cap T_{ai}$ comprises all those
constructs that are both changed in the fix commits of $j$ and
executed in the context of application $a$ because of its use of library $i$.

\subsection{Static Assessment of Vulnerable Code}
\label{subsec:static}

In addition to the analysis of \emph{actual} executions (dynamic analysis), our approach uses static analysis to determine whether the vulnerable constructs are \emph{potentially} executable. Specifically, we use static analysis in two different flavors. First, we use it to \emph{complement} the results of the dynamic analysis, by identifying the library constructs reachable from the application. Second, (Section~\ref{subsec:combination}) we \emph{combine} the two techniques by using the results of the dynamic analysis as input for the static analysis, thereby overcoming limitations of both techniques: static analyzers are known to struggle with dynamic code (such as, in Java, code loaded through reflection~\cite{DBLP:conf/icse/LandmanSV17}); on the other hand, dynamic (test-based) methods suffer from inadequate coverage of the possible execution paths.


Figure~\ref{fig:concept-a2c} illustrates how we use static analysis to complement the results of dynamic analysis. $R_{ai}$ represents the set of all constructs, either part of application $a$ or its bundled library $i$, that are found reachable starting from the application $a$ and thus can be potentially executed. Static analysis is performed by using a static analyzer (e.g., ~\cite{wala}) to compute a graph of all library constructs reachable from the application constructs. The intersection $C_j\cap R_{ai}$ comprises all
constructs that are both changed in the fix commit of $j$ and can be potentially executed.

\subsection{Combination of Dynamic and Static Assessment}
\label{subsec:combination}

Figure~\ref{fig:concept-t2c} illustrates how we combine static and dynamic analysis. In this case the set of constructs actually executed, $T_{ai}$, is used as starting point for the static analysis. The result is the set $R_{T_{ai}}$ of constructs reachable starting from the ones executed during the dynamic analysis. The intersection $C_{j}~\cap~R_{T_{ai}}$ comprises all
constructs that are both changed in the fix commit of $j$ and can be potentially executed.

\begin{figure}[t!]
\centering
\includegraphics[width=\columnwidth]{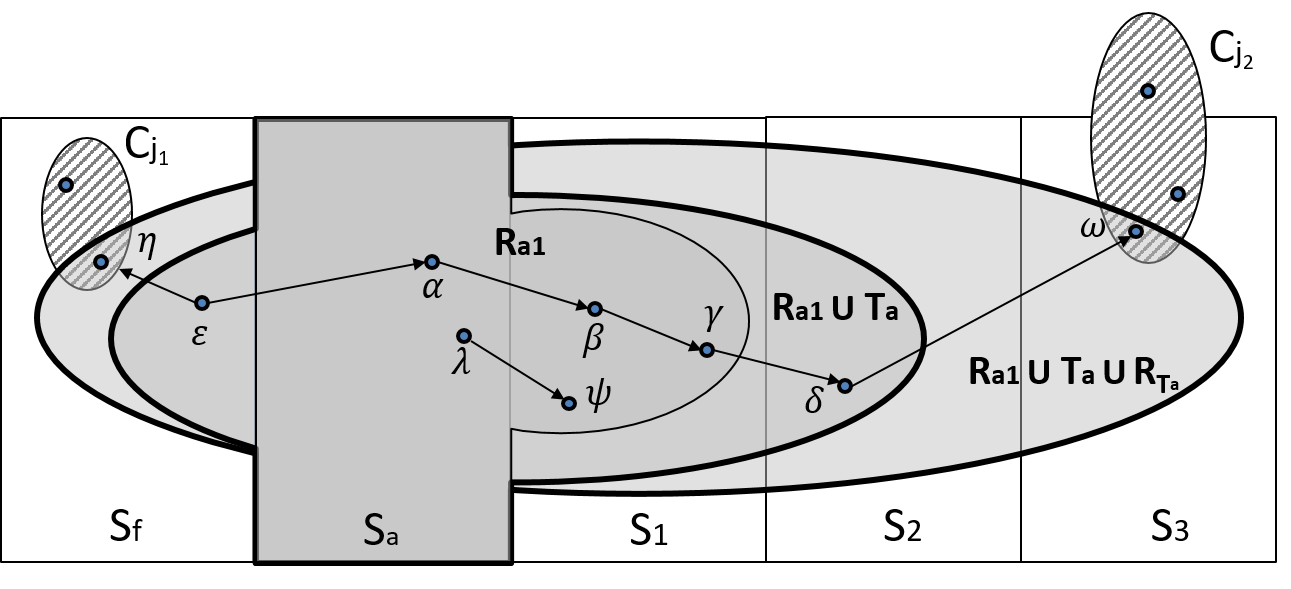}
\caption{Vulnerability analysis example}
\squeeze{4}
\label{fig:example}
\end{figure}

We explain the benefits of the combinations of the two techniques through the example in Figure~\ref{fig:example}.

In the following, we denote a library bundled within a software program with the term \emph{dependency}.
Let $S_a$ be a Java application having two direct dependencies $S_1$ and $S_f$ where $S_1$ has a direct dependency $S_2$ that in turn has a direct dependency $S_3$ (thereby $S_2$ and $S_3$ are transitive dependencies for the application $S_a$).  $S_1$ is a library offering a set of functionalities to be used by the application (e.g., Apache Commons FileUpload~\cite{fileupload}). Moreover the construct $\gamma$ of $S_1$ calls the construct $\delta$ of $S_2$ dynamically, e.g., by using Java reflection, which means the construct to be called is not known at compile time. $S_f$ is what we call a ``framework'' providing a skeleton whose functionalities are meant to call the application defining the specific operations (e.g., Apache Struts~\cite{struts}, Spring Framework~\cite{spring}). The key difference is the so-called \emph{inversion of control} as frameworks call the application instead of the other way round.

With the vulnerability detection step of Section~\ref{subsec:id}, our approach determines that $S_a$ includes vulnerable constructs for vulnerabilities $j_1$ and $j_2$ via the dependencies $S_f$ and $S_3$, respectively. Note that even if $S_3$ only contains two out of the three constructs of $C_{j_2}$, our approach is still able to detect the vulnerability.

We start the vulnerability analysis by running the static analysis of Section~\ref{subsec:static} that looks for all constructs potentially reachable from the constructs of $S_a$. The result is the set $R_{a1}$ including all constructs of $S_a$ and all constructs of $S_1$ reachable from $S_a$. As expected, $S_f$ is not reachable in this case as frameworks are not called by the application. Moreover, it is well known that static analysis cannot always identify dynamic calls like those performed using Java reflection. As the call from $\gamma$ to $\delta$ uses Java reflection, in this example only $S_1$ is statically reachable from the application. As shown in Figure~\ref{fig:example} $R_{a1}$ does not intersect with any of the vulnerable constructs.

The dynamic analysis of Figure~\ref{fig:concept-test} produces the set $T_a$ (omitted from the figure) of constructs that are actually executed. Though no intersection with the vulnerable constructs is found, the dynamic analysis increases the set of reachable constructs ($R_{a1}\cup T_a$ in Figure~\ref{fig:example}). In particular it complements static analysis revealing paths that static analysis missed. First, it contains construct $\epsilon$ of framework $S_f$ that calls construct $\alpha$ of the application. Second, it follows the dynamic call from~$\gamma$~to~$\delta$.

Combining static and dynamic analysis, as shown in Figure~\ref{fig:concept-t2c}, we can use static analysis with the constructs in $T_a$ as starting point. The result is the set $R_{T_a}$ (omitted in the figure) of all constructs that can be potentially executed starting from those actually executed~$T_a$.

After running all the analyses, we obtain the overall set $R_{a1} \cup T_a \cup R_{T_a}$ (shown with solid fill in Figure~\ref{fig:example}) of all constructs found reachable by at least one technique. Its intersection with $C_{j1}$ and $C_{j2}$ reveals that both vulnerabilities $j_1$ and $j_2$ are reachable, since one vulnerable construct for each of them is found in the intersection and is thus reachable ($\eta \in C_{j_1}$ is reachable from $\epsilon$ and $\omega \in C_{j_2}$ is reachable from~$\delta$).




\section{Vulnerability Mitigation}
\label{sec:metrics}
The analysis presented in Sections~\ref{subsec:dynamic} to \ref{subsec:combination} provides in-depth information about the control-flow between the code of the application and its dependencies. In the following, this information is leveraged to support application developers in mitigating vulnerable dependencies.

As long as non-vulnerable library versions are available, updating to one of those is the preferred solution to fix vulnerable application dependencies. And since the approach described in Section~\ref{subsec:background} depends on the presence of at least one fix commit, a non-vulnerable library version becomes available when the respective open-source project releases a version including this commit. However, it is well known that developers are reluctant to update dependencies because of the risk of breaking changes, the difficulties in understanding the implications of changes, and the overall migration effort \cite{DBLP:journals/ese/KulaGOII18,DBLP:conf/issta/MostafaRW17}.
Such risk and effort depend on the usage the application makes of the library, and on the amount of changes between the library version currently in use and the respective non-vulnerable version. As a result of the analysis described in Sections~\ref{subsec:dynamic} to \ref{subsec:combination}, the reachable share of each library
is known. Whether a construct with a given identifier is also available in other versions of a library can \review{always} be determined, for instance, by comparing compiled code with tools such as Dependency Finder~\cite{jeant}.
Among all the reachable constructs, of particular importance in the scope of mitigation are those which are called directly from the application,
as they provide a measure of how many times the application developer explicitly used the library. We define a \emph{touch point} as a pair of constructs $(c_1,c_2)$ such that $c_1 \in S_a$ is an application construct, $c_2 \in S_i$ is a library construct, and there exists a call from $c_1$ to $c_2$.
We define \emph{callee} the library construct called directly from the application, i.e., $c_2$.
In the example of Figure~\ref{fig:example} there are two touch points: $(\alpha, \beta)$ and $(\lambda, \psi)$, with $\beta$ and $\psi$ being the callees.

Given a library in use $S_i$ and its candidate replacement $S_j$, we define the following update metrics.

\noindent{\bf Callee Stability ($\mathit{CS}$). } Let $c_k^{(S_i)}$ with $k=1,\ldots,n$ be the callees of $S_i$, and $c_k^{(S_j)}=1$ if $c_k^{(S_i)} \in S_j$, $0$ otherwise. Then the callee stability is the number of callees of $S_i$ that exist in $S_j$ over the number of callees of $S_i$:

$$\mathit{CS} = \displaystyle\sum_{k=1}^{n} c_k^{(S_j)}/|\{ c_1^{(S_i)},\ldots,c_n^{(S_i)}\}| = \displaystyle\sum_{k=1}^{n} c_k^{(S_j)}/ n$$

If $S_j$ contains all the callees of $S_i$, then 
the callee stability is~1, to indicate that the constructs of $S_i$ called by the application exist also in library $S_j$. In case $S_j$ does not contain all the callees of $S_i$, then 
the callee stability is smaller than $1$ and reaches $0$ when none of the callees of $S_i$ is present in $S_j$.

\noindent{\bf Development Effort ($\mathit{DE}$). }
Let $a_k^{(S_i)}$ with $k=1,\ldots,n$ be the calls from the application to the callees of $S_i$, and $a_k^{(S_j)}=1$ if $a_k^{(S_i)} \not\in S_j$, $0$ otherwise.
The development effort for updating from library $S_i$ in use by the application to library $S_j$ is defined as the number of application calls that require a modification due to callees of $S_i$ that do not exist in $S_j$.

$$\mathit{DE} = \displaystyle\sum_{k=1}^{n} a_k^{(S_j)}$$

Compared to the callee stability, the development effort keeps into account the fact that each callee can be called multiple times within an application.

In Figure~\ref{fig:example}, for instance, each callee is called only once by $\alpha$ and $\lambda$ respectively. However, assuming that $\beta$ is called by two application constructs in addition to $\alpha$, and that it is not contained in the new library $S_j$, $\mathit{CS}=1/2$ whereas the $\mathit{DE}=3$. This reflects the fact that multiple calls need to be modified as a result of a change in a single callee.
\review{As defined, the development effort does not take into account the complexity of each modification but rather focuses on the number of modification required by the application as each one comes at the cost not only of updating the code (which could be automated to some extent) but also of testing it.}

\noindent{\bf Reachable Body Stability ($\mathit{RBS}$). } The reachable body stability is calculated in the same way as the callee stability, but instead of callees, it considers the reachable share of a library, i.e., the set of dynamically and/or statically reachable library constructs. Given the total number of constructs of $S_i$ reachable from the application, it measures the ratio of those that are contained as-is, i.e., with identical identifier and byte code, also in $S_j$. By quantifying the share of modified reachable constructs, this metric provides the likelihood that the behavior of the
library changes from $S_i$ to $S_j$. In case all reachable constructs of $S_i$ exist in $S_j$, then $\mathit{RBS}=1$ and thus there is a higher likelihood that the library change does not break the application.

\noindent{\bf Overall Body Stability ($\mathit{OBS}$). } The overall body stability is calculated similarly to $\mathit{RBS}$ but now considers all the constructs of $S_i$.
This metric provides the same indication as the one above but, by considering the entire library rather than only its reachable share, it is independent of the application-specific usage.

The above metrics support the application developer in estimating the effort and risks of updating a library. When several non-vulnerable libraries exist that are newer than the one  in use, they are all candidate replacements. By quantifying the changes to be performed on the application and the changes that the library underwent, our update metrics allow the developer to take an informed decision.

Note that the callee stability and development effort metrics only apply for dependencies that are called directly from the application, whereas the reachable and overall body stability also apply for transitive dependencies and frameworks.





\section{Evaluation}
\label{sec:evaluation}

The implementation of our approach for Java, has been successfully used at \SAP to perform over \vulasnumberofscansapprox~scans of about \vulasprojectcountapprox~applications since December 2016.
In the following, we illustrate how our approach works in a typical scan, applying it to a \SAP-internal web application that we adapted, for illustrative purposes, to include vulnerable OSS. The application allows users to upload files, such as documents or compressed archives, through an HTML form, inspects the file content and displays a summary to the user. It is developed using Maven~\cite{maven}, and depends on popular open-source libraries from the Apache Software Foundation, such as \code{Struts 2.3.24} (released on 3 May 2015), \code{Commons FileUpload 1.3.1} (6 February 2014), \code{POI 3.14} (6 March 2016) and \code{HttpClient 4.5.2} (21 February 2016). Overall, the application has 12 direct and 15 transitive compile-time dependencies.


The analysis is performed using an implementation of the approach described previously: 
Sections~\ref{subsec:id},~\ref{subsec:static}, and~\ref{subsec:combination} are implemented as \emph{goals} of a Maven plugin; the collection of traces during the dynamic analysis (Section~\ref{subsec:dynamic}) is performed by instrumenting all classes of both the application and all its dependencies as described in~\cite{icsme2015}. This happens either at runtime, when classes are loaded, or by modifying the byte code of the application before deploying it in a runtime environment such as Apache Tomcat.

\subsection{Detection and Analysis}

To illustrate the benefits of our approach,
we go through the analysis steps and highlight selected findings.
To demonstrate the added value of static analysis compared to \cite{icsme2015}, we perform it after dynamic analysis. However, our implementation also supports changing their order (or executing only a subset of the steps).

\noindent {\bf (1) Vulnerability Detection.} The first step is to create a \emph{bill of materials} (BOM), consisting of the constructs of the application and \emph{all its dependencies}, as explained in Section~\ref{subsec:id}.

\begin{figure*}
\centering
\includegraphics[width=\textwidth]{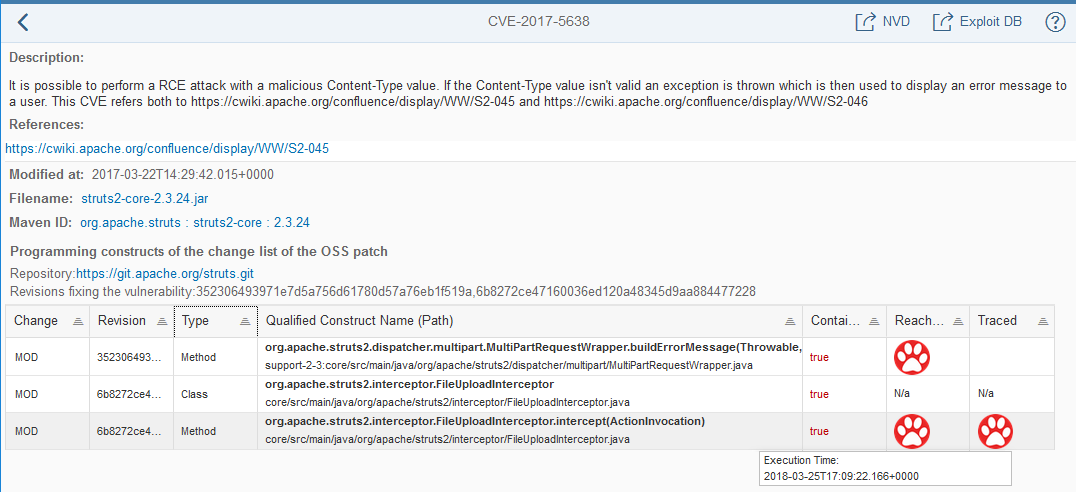}
\caption{Static analysis starting from traced methods provides further evidence that \cve{CVE-2017-5638} is relevant}
\label{fig:screenshot-t2c}
\end{figure*}

Vulnerabilities in a library are detected by intersecting the set of constructs found in (the BOM of) that library with the vulnerable constructs of all the vulnerabilities known to our knowledge base. As an example, the bottom part of Figure~\ref{fig:screenshot-t2c} shows a table listing the vulnerable constructs for \cve{CVE-2017-5638} (columns \emph{Type} and \emph{Qualified Construct Name}) together with the respective change operation (column \emph{Change}), as well as the information that those constructs are actually present in the Java archive corresponding to \code{struts-core:2.3.24} (column \emph{Contained}).

The vulnerability detection step reveals that our sample application includes vulnerable code related to 25 different vulnerabilities, affecting nine different compile-time dependencies: seven are \emph{direct}, while the remaining two (\code{ognl:3.0.6} and \code{xwork-core:2.3.24}\footnote{Maven dependencies are denoted using their artifact identifier followed by, where necessary, a colon and their version. Group identifiers are omitted for brevity.}, pulled in through \code{struts2-core:2.3.24}) are transitive.



\begin{figure}
\centering
\includegraphics[width=\columnwidth]{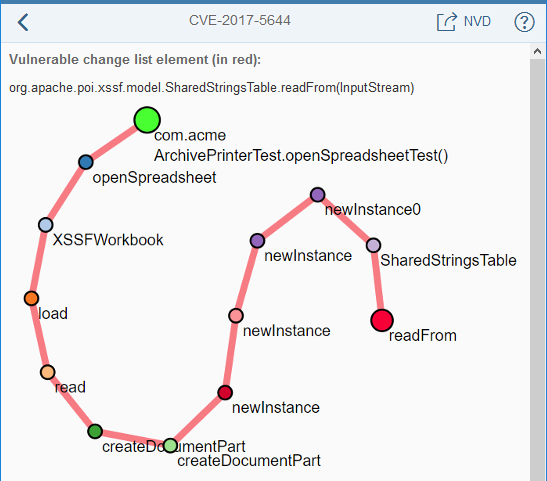}
\caption{Unit tests reveal the execution of vulnerable construct \code{SharedStringsTable.readFrom(InputStream)} in \code{poi-ooxml}}
\label{fig:screenshot-a2c}
\end{figure}

\noindent {\bf (2) Dynamic Assessment (Unit tests).} The execution of unit tests reveals that vulnerable constructs related to three vulnerabilities are executed, e.g., the method \code{URIBuilder.normalizePath(String)}\footnote{Where possible, Java package and class names are omitted for brevity.}, which is part of \code{httpclient:4.5.2} and subject to vulnerability \cve{HTTPCLIENT-1803}~\cite{httpclient-1803}.
%
Another example is shown in Figure~\ref{fig:screenshot-a2c}: method \code{SharedStringsTable.readFrom(InputStream)}, which is part of \code{poi-ooxml:3.14} and subject to \cve{CVE-2017-5644}, is invoked in the context of unit test \code{openSpreadsheetTest}.
The fact that reflection is heavily used inside the \code{poi-ooxml} method \code{XSSFFactory.createDocumentPart(Class,Class[],Object[])}
(as visible from a sequence of four invocations of \code{newInstance}, see figure) makes it difficult for static analysis to determine the reachability of the vulnerable method.

\begin{figure}
\centering
\includegraphics[width=\columnwidth]{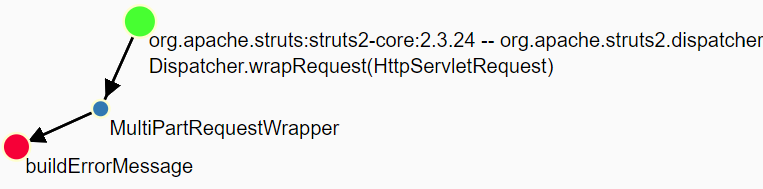}
\caption{Combined analysis reveals the reachability of \code{MultiPartRequestWrapper.buildErrorMessage(\ldots)} in \code{struts2-core}}
\label{fig:screenshot-t2c-path}
\end{figure}

\noindent {\bf (3) Dynamic Assessment (Integration tests).} The execution of integration tests is done using an instrumented version of the application deployed in a runtime container. They reveal the execution of vulnerable code related to eight additional vulnerabilities, all affecting \code{struts2-core}, or its dependencies \code{ognl} and \code{xwork-core}.
As an example, the last line of the table in Figure~\ref{fig:screenshot-t2c} shows a vulnerable construct of \cve{CVE-2017-5638}, \code{FileUploadInterceptor.intercept(ActionInvocation)}, whose actual execution is traced (column \emph{Traced}) at the reported time.
This method is included in \code{struts2-core 2.3.24} which is part of the \code{Struts2} framework and exemplifies the inversion of control (IoC) happening when frameworks invoke application code.

\noindent {\bf (4) Static Assessment.} The static analysis starting from application constructs reveals that the constructor \code{MultipartStream(InputStream,byte[],int,ProgressNotifier)}, part of \code{commons-fileupload:1.3.1} and subject to \cve{CVE-2016-3092}, is reachable from the application.
Dynamic analysis was not able to trace its execution due to the limited test coverage.


On the other hand, static analysis starting from the application constructs falls short in the presence of IoC. As application methods are called \emph{by the framework}, there is no path on the call graph starting from application and reaching framework constructs that are involved in the IoC mechanism.


\noindent {\bf (5) Combination of Static and Dynamic Assessment.} The static analysis starting from constructs traced with dynamic analysis provides additional evidence regarding the relevance of \cve{CVE-2017-5638} (the vulnerability that was exploited in the Equifax breach~\cite{equifax}). In addition to the execution of method \code{FileUploadInterceptor.intercept(ActionInvocation)} during step~3, the combination of static and dynamic analysis reveals that method
\code{MultiPartRequestWrapper.buildErrorMessage(Throwable, Object[])}, included in \code{struts2-core 2.3.24}, is reachable with two calls from the traced method \code{Dispatcher.wrapRequest(HttpServletRequest)}, as shown in Figure~\ref{fig:screenshot-t2c-path}. Its reachability is indicated with the red paw icons in the table containing the construct changes for \cve{CVE-2017-5638} (cf. the two right-most columns of the table in Figure~\ref{fig:screenshot-t2c}).

\note{SP: check the numbers}
The {\it value of combining the two analysis techniques} becomes more evident when considering all applications scanned with our approach:
vulnerable constructs are reachable, statically or dynamically, in 131 out of \vulasprojectcount applications.
In particular we observed 390 pairs of applications and vulnerabilities whose constructs were reachable. In 32 cases, the reachability could only be determined through the combination of techniques, which represents a 7.9\% increase of evidence that vulnerable code is potentially executable.

\subsection{Mitigation}

During the execution of dynamic analysis (steps~2 and~3) and static analysis (steps~4 and~5), touch points and reachable constructs are collected. They are the basis for the computation of the update metrics for the application at hand.

\begin{figure*}
\centering
\includegraphics[width=\textwidth]{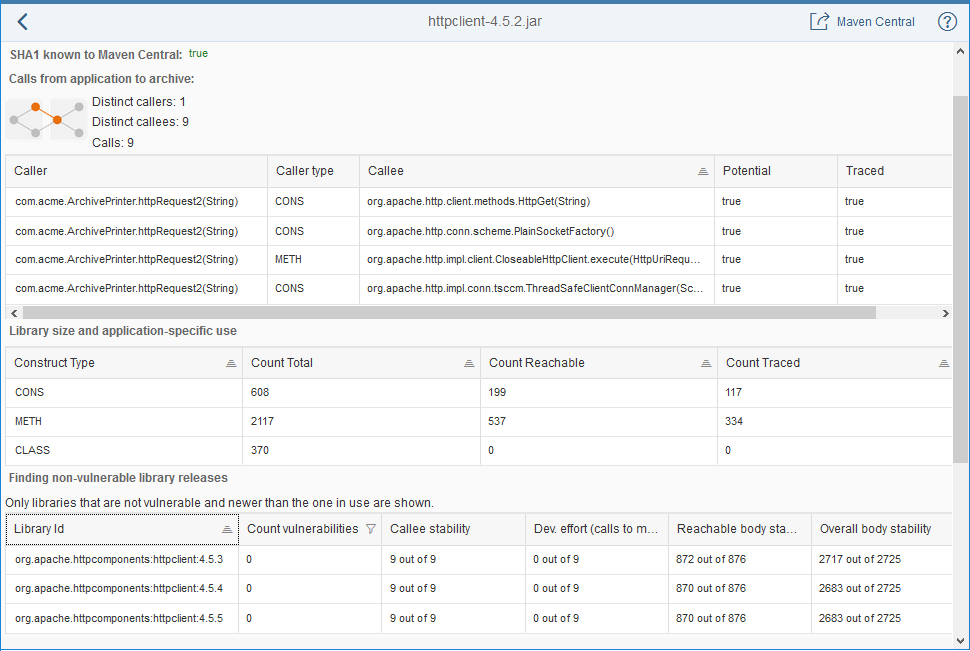}
\caption{Touch points, reachable constructs and update metrics for \code{httpclient:4.5.2}}
\label{fig:screenshot-mitigation}
\end{figure*}

Figure~\ref{fig:screenshot-mitigation} shows that for \code{httpclient:4.5.2}, one of the direct dependencies of the application, there are nine touch points between the application and the library. The application method \code{ArchivePrinter.httpRequest2(String)}, for instance, calls the constructor \code{HttpGet(String)} (cf. first table in the figure). This invocation was observed during dynamic analysis, and was also found by static analysis (cf. rightmost columns in the first table in the figure).
The second table of the figure shows the number of constructs of \code{httpclient:4.5.2} by type. For example, of the  608 constructors (\code{CONS}), 199 were found reachable by static analysis, and 117 were actually executed during  tests.

The table at the bottom of Figure~\ref{fig:screenshot-mitigation} shows the
update metrics that can guide the developer in the selection of a non-vulnerable replacement for \code{httpclient:4.5.2}. Each table row corresponds to a release of \code{Apache HttpClient} that is not subject to any vulnerability known to our knowledge base, hence, the developer is advised to choose among the three versions: \code{4.5.3}, \code{4.5.4} and \code{4.5.5}. The callees of all touch points exist in all of those versions, hence, the update to any of those would not result in signature incompatibilities (cf. columns 3 and 4). The $\mathit{RBS}$ metric indicates that 872 out of 876 reachable constructs of type method and constructor
are also present in release \code{4.5.3} (870 out of 876 in \code{4.5.4} and \code{4.5.5}). The $\mathit{OBS}$ metric is also  relatively high for all three non-vulnerable releases, thus, the developer would likely choose \code{httpclient:4.5.5} in order to update the vulnerable library.

\begin{figure}
\centering
\includegraphics[width=\columnwidth]{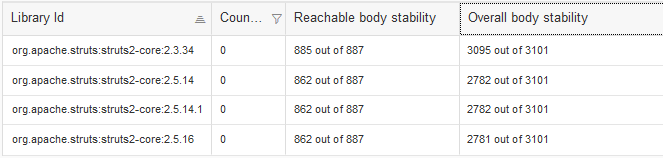}
\caption{Update metrics for \code{struts2-core:2.3.24}}
\label{fig:screenshot-mitigation-struts}
\end{figure}

While the update decision is relatively straightforward for \code{httpclient}, it is more difficult for \code{strut2-core}, since there are non-vulnerable replacements from both the 2.3 and the 2.5 branch (cf. Figure~\ref{fig:screenshot-mitigation-struts}). Here, the $\mathit{RBS}$ and $\mathit{OBS}$ metrics indicate a more significant change of constructs between the current version \code{struts-core:2.3.24} and the latest version of the 2.5 branch ($\mathit{RBS}$=862/887 and $\mathit{OBS}$=2781/3101) than between the current version and the latest version of the 2.3 branch ($\mathit{RBS}$=885/887 and $\mathit{OBS}$=3095/3101). Hence, the developer may be more inclined to stick to the 2.3 branch, thus updating to \code{struts-core:2.3.34} rather than to \code{struts2-core:2.5.16}.

\section{Experience Report}
\label{sec:experience}


\subsection{From Research Prototype to Industrial-Grade Solution.}


The approach presented in this paper is implemented as a \SAP-internal tool called \VULAS. Initially, a research prototype was used to run pilots with a small number of development units, to clarify the needs of real-life development projects and to evaluate the viability of our approach.

From the feedback gathered from the users of the prototype, we could make two clear observations:

\paragraph{Precision} Developers feared that using \emph{yet another tool} (general static code analyzers as well as dynamic security testing tools were widely available to scan the code of open-source dependencies used in \SAP's products) would mean being flooded with more findings referring to \emph{potential} issues, many of which  irrelevant (not exploitable) in practice. Especially the more security-aware among developers were reluctant to use metadata-based tools, because of their excessive rate of false-positives and false-negatives. As a matter of fact, frequent false-positives can challenge the adoption of a tool as much as false-negatives do. This observation confirmed the importance of \review{our decision to strive for} a \emph{reliable, precise} method to detect the \emph{actual presence} of vulnerable \emph{code} in a given library.

\paragraph{Inobtrusiveness and Automation} A tool whose adoption requires changes in the development practices is extremely hard to promote. We made the choice to integrate with the de-facto standard build processes and tools, making \VULAS a pluggable element of the automated build tool-chain that could be enabled with minimal configuration effort. Automating vulnerability scans has the additional benefit that issues are detected in a more timely manner, allowing better prioritization and effort allocation to identify and apply cost-effective fixes.

After the prototyping phase, as the demand for \VULAS started to increase, the tool underwent a major reimplementation
in order to make it more flexible and scalable, by adopting a micro-services architecture and deploying it onto \SAP{}'s internal cloud infrastructure.

With the growth of the user base, we could observe that the tool is used differently depending on the phase of the development life-cycle.

Projects in the earliest phases of development, and particularly those that have not yet been released to customers, are mainly concerned with identifying as early as possible the dependency on a vulnerable library. The large majority of projects (roughly 90\%) \review{are scanned with} \VULAS routinely, as part of an automated build pipeline, in which vulnerability detection (based purely on the analysis of the constructs of the application and its libraries as presented in Section~\ref{subsec:id}) is performed at each commit, whereas deeper analysis is only run as a nightly (or weekly) job. The performance offered by the vulnerability detection implemented in \VULAS is adequate for frequent scans, since its average execution time is about 70 seconds (static and dynamic analysis can take hours to complete, depending on the size of the application).

We observed that in this phase, the resistance to updates (especially, to minor releases) is quite low, and developers tend to update their OSS dependencies in a short time-frame.

Deeper analysis of the vulnerability becomes more and more critical the closer the project gets to the date of release to customers, and stays so throughout the operational lifetime of the application because new vulnerabilities impacting the application could be discovered at any point in time after the release.
After the release date, \VULAS is used mostly in manual scans, as a \emph{program comprehension} tool, to achieve a deeper understanding of whether and how vulnerable code is reachable in a target application, what its concrete impact is, and what remediation options are available.

To  deal also with legacy applications, built without modern tools such as Maven~\cite{maven} or Gradle~\cite{gradle}, a dedicated command-line version of \VULAS is often used.
%
%

We found that both in new and legacy applications, library artifacts are very often renamed in an ad-hoc, often inconsistent manner, and the content of one or more original libraries might be extracted and repackaged as a single self-contained archive. In this context, metadata is most often incomplete or missing, which makes our \emph{code-centric} approach critical. The method implemented in \VULAS, whereby vulnerable code is identified in the (byte)code of libraries, does not suffer from the limitations of the approaches based on metadata (such as OWASP Dependency Check~\cite{owasp-dc}, which relies on Maven artifact identifiers, file names or the content of manifest files to work correctly).

\begin{figure}
\centering
\includegraphics[width=\columnwidth]{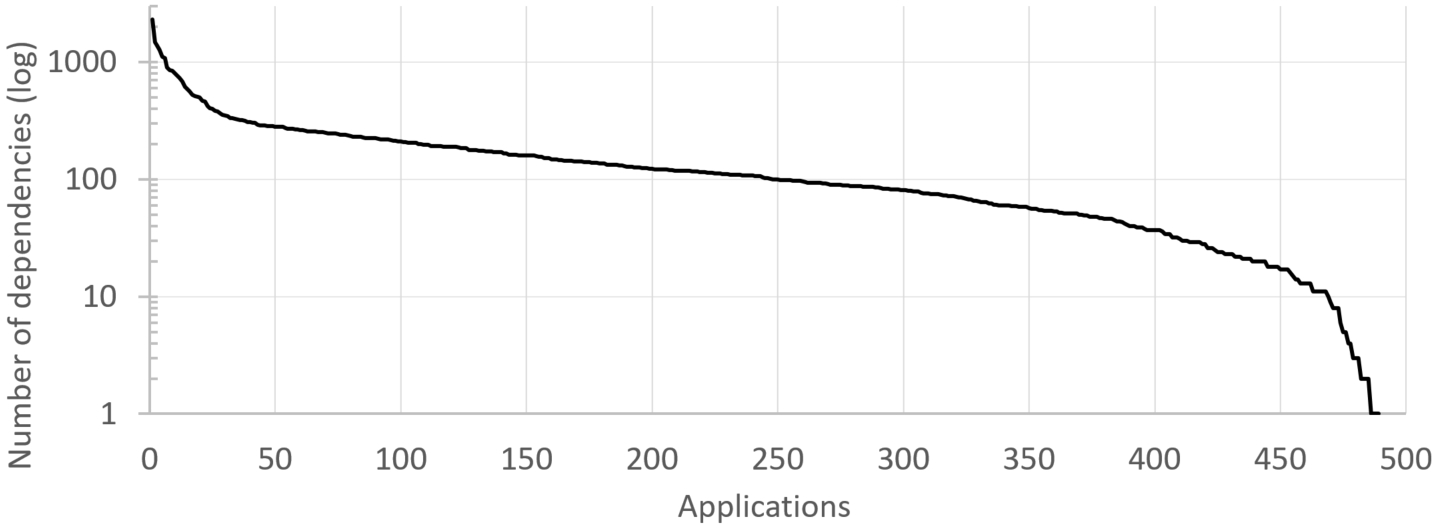}
\caption{Number of dependencies in the applications analyzed.}
\label{fig:APP-DEPS}
\end{figure}

At the time of writing, \VULAS is the officially recommended tool at \SAP to scan Java projects. It is used by over \vulasprojectcountapprox distinct development projects, and its adoption has been growing at a rate of 15 to 20 new applications per week since the beginning of 2018.
The sizes of the applications scanned as of today are plotted in Figure~\ref{fig:APP-DEPS}.
During these scans, we detected about 30000 pairs of vulnerabilities and libraries. More than a half of these pairs concern libraries that are not published to Maven central (because they have been recompiled, repackaged, or otherwise manipulated).

Despite its success, several challenges, pertaining to either organizational or technical aspects, remain to be addressed. In the remainder of this section we briefly discuss the most important ones.

\subsection{Challenges}

\subsubsection{Developer Opt-in vs. Central Scans}
To foster the uptake by developers, one has to minimize the required changes of development artifacts and processes. Plug-ins for common build tools support this goal, and the provision of templates for continuous integration pipelines goes as far as enabling in-depth \VULAS scans by means of boolean flags. Still, the tool adoption ultimately depends on the developer's initiative, even at that degree of integration and automation. Awareness campaigns, trainings and other organizational measures are one means to this end, however, they require significant resources in large, international and heterogeneous development organizations. Future work aims at overcoming such issues by running fully-automated scans at a few central elements of an organization's development infrastructure, e.g., its source code or artifact repositories.

\subsubsection{Decision-Support vs. Decision-Making}
\VULAS is essentially a \emph{fact-finding} tool, whose goal is to provide comprehensive evidence that vulnerable code is \emph{included} and \emph{reachable} in a given application. Clearly, it is cannot prove (decide) whether vulnerable code can or cannot be \emph{exploited}. However, we occasionally observe this expectation, particularly when library updates are difficult and expensive. Again, trainings and security-awareness campaigns are essential to ensure that developers internalize that they are responsible for drawing the final conclusion in regards to the exploitability of each vulnerability, especially before sticking to a vulnerable open-source version. Also, while such conclusions and the decision to upgrade (or not) a library have to be documented, we decided to keep this functionality out of the scope of \VULAS and to stick to its fact-finding mission.


\subsubsection{Vulnerability Knowledge-Base}
\label{subsubsec:db}
The creation and maintenance of a comprehensive vulnerability database is key for our approach. However, the required vulnerability information is scattered across many different sources, e.g., public vulnerability databases, issue trackers and security advisories of individual projects, or source code repositories. Moreover, as discussed in~\cite{icsme2015,alqahtani2016tracing}, different sources like the National Vulnerability Database and code repositories are difficult to integrate using the existing data. \review{To evaluate the completeness of the our current knowledge base we conducted an internal study of the vulnerability dataset, which concluded that it covers 90\% of all NVD vulnerabilities reported for Java open-source projects.} At the time of writing, we are exploring the use of machine-learning methods to automate the identification of fix commits and to ease the maintenance of a rich knowledge base where vulnerabilities are linked to the corresponding source code changes. The maintenance of such a detailed knowledge base,
which is done manually as of today, would greatly benefit from a coordinated approach to vulnerability disclosure and patch release across the open-source community, and through the governance exercised by established institutions, such as the Apache Software Foundation. 





\subsubsection{Shallow vs. Deep Updates} 
The metrics of Section~\ref{sec:metrics} support the update to non-vulnerable library versions. In the case of transitive dependencies, however, developers are required to interfere with the transparency and automation of the dependency resolution mechanism. In fact, one would need to add the updated non-vulnerable version of the library as a direct dependency, thereby taking the risk of future incompatibilities. 
Therefore, the mitigation strategy could be optimized to recommend the update of the library that is ``nearest'' to the application. As an example, to mitigate the vulnerable library $S_3$ of Figure~\ref{fig:example}, a newer version of $S_1$ can be recommended if it pulls in a fixed version of~$S_3$.


\subsubsection{Problematic Types of Vulnerabilities} 
By identifying a vulnerable dependency through the presence of vulnerable code, our approach is robust against false-positives and false-negatives, as typical for solutions based on the mapping of library and vulnerability metadata. As a drawback, the fraction of vulnerabilities whose fix does not involve any code change, e.g., those that are fixed by modifying a default configuration, cannot be covered by our code-centric approach described in Section~\ref{subsec:background}. Nevertheless, \VULAS is able to cover such cases, which are relatively rare compared to code-related vulnerabilities, by flagging entire libraries as affected.

By assessing the exploitability of a vulnerability in terms of potential or actual code execution, our approach provides evidence about the application-specific usage of vulnerable code. However, this approach does not work with vulnerabilities that are due to the deserialization of untrusted data, where the  mere presence of so-called \emph{deserialization gadgets} in the application \emph{classpath} can cause the application to be vulnerable (regardless of whether they can be reached during normal program execution). Attackers exploit the behavior of the Java serialization mechanism, which creates objects (through deserialization) as long as the definition of the respective class is known, regardless of whether the application actively uses it or not.

\subsubsection{Shortcomings of Name-based Construct Identification}

While the approach relies on the generic concept of construct identifiers, \VULAS uses Java \emph{fully-qualified names}
in its implementation.
This implementation choice supports other programming languages as long as they have a comparable naming scheme, and its development community follows consistent naming conventions (that is, construct names can be assumed to be globally unique, as in Java). However, this property is not satisfied for certain languages, hence, the construct identification cannot be done just by using their fully-qualified names, but must consider other information, such as their bodies.







\section{Related work}
\label{sec:rel-work}
There exist several free~\cite{owasp-dc} and commercial tools~\cite{whitesource,snyk,blackduck,sourceclear} for detecting vulnerabilities in OSS components.
\review{In~\cite{icsme2015} we showed that our approach outperforms state-of-the-art tools \emph{with respect to vulnerability detection}.}
Though \cite{sourceclear} claims to perform static analysis to eliminate false positives, there is no public description of their approach available. OWASP Dependency Check~\cite{owasp-dc} is used in~\cite{cadariu2015tracking} to create a vulnerability alert service and to perform an empirical investigation about the usage of vulnerable components in proprietary software. The results showed that 54 out of 75 of the projects analyzed have at least one vulnerable library. However the results had to be manually reviewed, as the matching of vulnerabilities to libraries showed low precision.
Alqahtani et al. proposed an ontology-based approach to establish a link between vulnerability databases and software repositories~\cite{alqahtani2016tracing}. The mapping resulting from their approach
yields a precision that is 5\% lower than OWASP Dependency Check.
All these  approaches and tools differ from ours in that they focus on vulnerability detection based on metadata, and do not provide application-specific reachability assessment nor mitigation proposals.

\review{Our previous work~\cite{icsme2015} already goes} beyond the detection of a vulnerability by performing reachability analysis:
\review{we used} dynamic analysis to establish whether vulnerable code is actually executed.
In this work, we extend \cite{icsme2015} including also static analysis and providing a novel combination of static and dynamic analysis.
To the best of our knowledge none of the existing works and tools combines static and dynamic analysis, nor provides application-specific mitigation proposals.

The screening test approach devised by Dashevskyi et al.~\cite{dashevskyi2018tse}
represents a scalable solution to the difficult problem of determining, at the time when a new vulnerability
is disclosed for a given OSS component, which other previous versions are also affected by the same vulnerability.
We tackle the same problem by comparing ASTs of the constructs involved in the vulnerability across the different
versions of the affected component; however, due to space constraints, the details of our method are not covered in this paper.

The empirical study conducted by Kula et al. on library migrations of 4600 GitHub projects showed that 81.5\% of them do not update their direct library dependencies, not even when they are affected by publicly known vulnerabilities~\cite{kula2017ese}. In particular, that study highlights the lack of awareness about security vulnerabilities.
Considering 147 Apache software projects, \cite{bavota2015ese} studied the evolution of dependencies and found that applications tend to update their dependencies to newer releases containing substantial changes.
Tools based on reward or incentives to trigger the update of out-of-date dependencies exist (e.g.,~\cite{david-dm,greenkeeper}), however as shown in~\cite{pull-requests}, project developers are mostly concerned about breaking changes and mechanisms are needed to provide--next to transparency--a motivation for the update and confidence measures to estimate the risk of performing the update. By automatically detecting vulnerabilities, providing evidence about the reachability of the vulnerable code, and supporting mitigation via update metrics, our work addresses the need of motivating updates and estimating  effort and risk.

In~\cite{visser2012} four metrics to measure the stability of libraries through time are proposed. In particular it considers the removal of units (constructs in our context), the amount of change  in existing constructs, the ratio of change in new and old constructs, and the percentage of new constructs.
Similar to our work, the metrics are meant to be representative for the amount of work required to update a certain library and thus they also consider usages of library methods in other projects. However the main focus of~\cite{visser2012} is the \emph{library}, and the metrics are used to measure its stability over time given a set of projects.
Though some of their metrics \emph{ingredients} can also be considered in our work, the metrics we propose are about the application-specific library usage. Moreover, our metrics benefit of our in-depth analysis of the application (e.g., some usages of the libraries that can only be observed with a combination of static and dynamic analysis).

Raemaekers et al. studied breaking changes in library releases over seven years and showed that they occur with the same frequency in major and minor releases~\cite{breakingChanges}. This shows that the rules of \emph{semantic versioning}, according to which breaking changes are only allowed in major releases, are not followed in practice. It also shows that top three most frequent breaking changes involve a deletion of methods, classes, fields, respectively. This result reinforces our belief that our update metrics based on measuring the removal of program constructs provides a critical information.


Mileva et al. studied the usage of different library
versions and provided a tool to suggest which one to use based on the choice of the majority of similar users~\cite{Mileva:2009}. Our work differs from theirs, in that we provide quantitative measures to support the user in selecting a non-vulnerable library.

Existing works on library migration~\cite{apiWave,semDiff2009,Nguyen:2010} are complementary to our approach in that they support developers in evolving their code to adapt to new libraries or library versions. \cite{apiWave} proposes a tool that keeps track of API popularity and migration of major frameworks/libraries, amounting to 650 Github projects resulting in 320000 APIs at the time of publication. 
\cite{semDiff2009} describes tools able to recommend replacements for framework methods accessed by a client program and deleted over time. \cite{Nguyen:2010} presents a tool able to recommend complex adaptations learned from already migrated clients or the library itself.

\section{Conclusion}
\label{sec:conclusion}

The unique contribution of this paper is the use of static analysis and its combination with dynamic analysis to support the application-specific assessment and mitigation of open-source vulnerabilities.
This approach further advances the code-centric detection and dynamic analysis of vulnerable dependencies \review{ we originally proposed in~\cite{icsme2015}}.

The accuracy and application-specific nature of our method improves over
state-of-the-art approaches, which commonly depend on metadata. \VULAS, the implementation of our approach for Java, was chosen by \SAP among several candidates as the recommended OSS vulnerability scanner. Since December 2016, it has been used for over \vulasnumberofscansapprox~scans of about \vulasprojectcountapprox~applications, which demonstrates the viability and scalability of the approach.

The variety of programming languages used in today's software systems pushes us to extend
\VULAS to support languages other than Java. However, fully-qualified names can be inadequate to uniquely identify the relevant program constructs in certain languages, so we are considering the use of information extracted from the construct bodies.

Finally, the problem of systematically linking open-source vulnerability information to the corresponding source code changes (the fix) remains open. Maintaining a comprehensive knowledge base of rich, detailed vulnerability data is critical to all vulnerability management approaches and requires considerable effort. While this effort could be substantially reduced creating specialized tools, we strongly believe that the maintenance of this knowledge base should become an industry-wide, coordinated effort, whose outcome would benefit the whole software industry.

\subsection*{Acknowledgements.}
We are also grateful to our colleagues Michele Bezzi, Luca Compagna, C\'edric Dangremont,
and Brian Duffy for their insightful comments on early drafts of this work.

%
%


\bibliographystyle{IEEEtran}
\bibliography{biblio}

\end{document}